\documentclass[lettersize,journal]{IEEEtran}
\usepackage{amsmath,amsfonts}
\usepackage{algorithmic}
\usepackage{array}
\usepackage[caption=false,font=normalsize,labelfont=sf,textfont=sf]{subfig}
\usepackage{textcomp}
\usepackage{stfloats}
\usepackage{url}
\usepackage{verbatim}
\usepackage{graphicx}
\usepackage{xcolor}
\usepackage{multirow}

\def\BibTeX{{\rm B\kern-.05em{\sc i\kern-.025em b}\kern-.08em
    T\kern-.1667em\lower.7ex\hbox{E}\kern-.125emX}}
\usepackage{balance}

\begin{document}
\title{{\fontsize{21}{0}\selectfont 
Metamobility: Connecting Future Mobility with Metaverse}}

\author{Haoxin~Wang,~\IEEEmembership{Member,~IEEE,}
        Ziran~Wang,~\IEEEmembership{Member,~IEEE,}
        Dawei~Chen,~\IEEEmembership{Member,~IEEE,}
        Qiang~Liu,~\IEEEmembership{Member,~IEEE,}
        Hongyu~Ke,       
        and Kyungtae~Han,~\IEEEmembership{Senior Member,~IEEE,}
        
\thanks{Haoxin Wang and Hongyu Ke are with Georgia State University, Department of Computer Science, 25 Park Place, Atlanta, GA 30303 (email: haoxinwang@gsu.edu; hk3@student.gsu.edu); 

Ziran Wang is with Purdue University, College of Engineering, 550 Stadium Mall Drive, West Lafayette, IN 47907 (email: ryanwang11@hotmail.com);

Dawei Chen and Kyungtae Han are with Toyota Motor North America R\&D, InfoTech Labs, 465 Bernardo Avenue, Mountain View, CA 94043 (e-mail: dawei.chen1@toyota.com; kyungtae.han@toyota.com); 

Qiang Liu is with University of Nebraska-Lincoln, School of Computing, 1144 T Street, Lincoln, NE 68588 (email: qiang.liu@unl.edu).
}}


\maketitle

\begin{abstract}
A Metaverse is a perpetual, immersive, and shared digital universe that is linked to but beyond the physical reality, and this emerging technology is attracting enormous attention from different industries. In this article, we define the first holistic realization of the metaverse in the mobility domain, coined as ``metamobility". We present our vision of what metamobility will be and describe its basic architecture. We also propose two use cases, tactile live maps and meta-empowered advanced driver-assistance systems (ADAS), to demonstrate how the metamobility will benefit and reshape future mobility systems. Each use case is discussed from the perspective of the technology evolution, future vision, and critical research challenges, respectively. Finally, we identify multiple concrete open research issues for the development and deployment of the metamobility.
\end{abstract}


\section{Background and Motivation}
\subsection{Future Mobility}

\begin{figure*}[t]
    \centering
    \includegraphics[width=\linewidth]{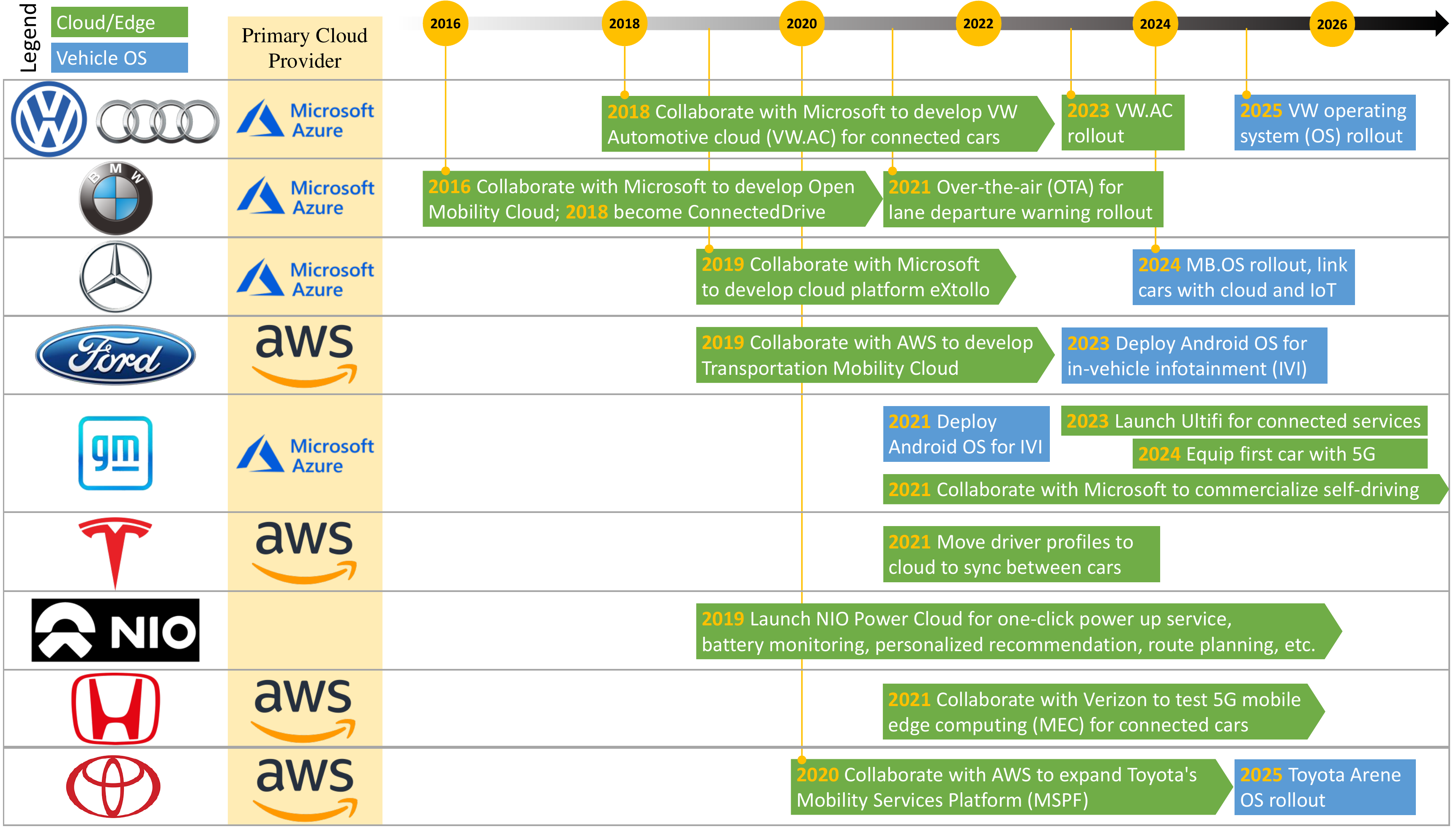}
    \caption{A timeline of selected OEMs' research activities and product/service roll-outs from 2016 to 2026 in terms of car connectivity.}
    \label{Fig:OEM_benchmark}
\end{figure*}

\IEEEPARstart{L}{ess} than half a decade, the expeditious evolution of wireless communications, Artificial Intelligence (AI), and high-performance computing (HPC), has been reinventing the mobility concept and systems. For example, in 2019, Toyota announced a profound transformation from being an automaker to becoming a mobility company, with an emphasis on \underline{C}onnectivity, \underline{A}utonomous driving, \underline{S}hared mobility, and \underline{E}lectrification of vehicles (CASE) \cite{ToyotaCASE}. Meanwhile, other major automotive original equipment manufacturers (OEMs), including Volkswagen, Audi, BMW, etc., are investing heavily in future mobility solutions to enhance their core competencies and to ingratiate themselves with customers. Although OEMs may frame their own blueprints for the future mobility, they share the same ambition of \textit{creating a mobility society in which safe, sustainable, frictionless, fun, and personalized transportation is universal by leveraging CASE technologies.} Fig. \ref{Fig:OEM_benchmark} illustrates a timeline of selected OEMs' research activities and product/service roll-outs in terms of connectivity, one of the cornerstones of future mobility. A trendy shift in OEM strategies is partnering with technology giants, such as Google, Amazon, and Microsoft, to develop cloud platforms, automotive operating systems (OS), AI, etc. 

\subsection{Metamobility}
\begin{figure*}[t]
    \centering
    \includegraphics[width=\linewidth]{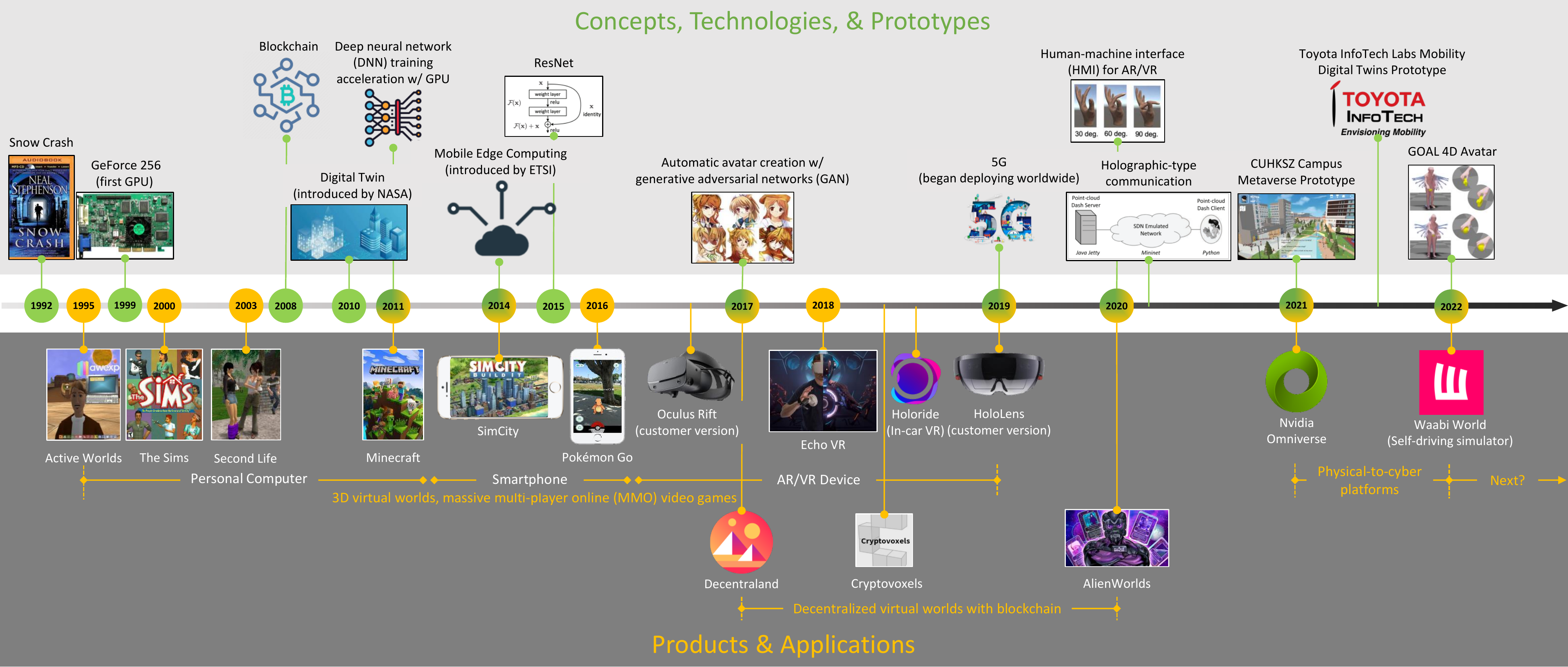}
    \caption{Evolution of the metaverse from 1992 to 2022, including the development of products \& applications and the expansion of concepts, technologies, \& prototypes (information source partially from \cite{lee2021all}); automatic avatar creation with GAN \cite{jin2017towards}; HMI for AR/VR \cite{zhu2020haptic}; holographic-type communication \cite{clemm2020toward}; CUHKSZ campus metaverse prototype \cite{duan2021metaverse}; Toyota InfoTech Labs mobility digital twin prototype \cite{9724183}; and GOAL 4D avatar \cite{taheri2022goal}.}
    \label{Fig:metaverse}
\end{figure*}

Metaverse, as an emerging representation of the immersive Internet, has attracted enormous attention in both academia and industry. Fig. \ref{Fig:metaverse} illustrates the timeline of the metaverse development from 1992 (the year that the term ``metaverse'' was coined in a science fiction named Snow Crash \cite{stephenson2003snow}) to 2022, including remarkable concepts, technologies, prototypes, products, and applications. Although the definitions of the metaverse vary in different versions \cite{duan2021metaverse,lee2021all,MetaHorizon,Omniverse}, the basic concepts are essentially the same: \textit{the metaverse is a perpetual, immersive, and shared digital universe (``verse'') that is linked to but beyond (``meta'') the physical reality}. Recently, several prominent prototypes and applications have been proposed by forerunners. The Chinese University of Hong Kong, Shenzhen (CUHKSZ) Metaverse \cite{duan2021metaverse}, for instance, is a campus metaverse prototype that allows students and faculty to perform immersive interactions in a real-virtual mixed world. Along with its rapid development in gaming \cite{MetaHorizon}, social media \cite{duan2021metaverse}, and manufacturing \cite{Omniverse}, the metaverse retains potential in future mobility as well.

Designing a metaverse solution for future mobility is strongly motivated by the ambition of achieving a frictionless, fun, and personalized mobility society. Additionally, the emergence of CASE technologies provides a catalyst for connecting the metaverse with future mobility. However, to the best of our knowledge, there has been no effort to discuss the confluence of the metaverse and future mobility technologies. In this article, a terminology, ``Metamobility'', is coined and defined as \textit{a holistic realization of the metaverse (``meta'') in the mobility domain (``mobility'') with the support of CASE technologies}. The metamobility is capable of driving customers in both physical and digital spaces, where connected and automated vehicles (CAVs) or other mobile entities such as urban air mobility (UAM) will be physical carriers for customers to access and interact with both real and virtual worlds. For example, during the COVID-19 pandemic, many international travels are suspended due to the pandemic prevention requirements. While the metamobility could provide great accessibility to allow people quarantining in Shanghai to enjoy an immersive Tokyo city tour in the cyber world by remotely interacting with a car (e.g., Toyota e-Palette self-driving car) in the physical world. Two essential features of the metamobility could be observed in the presented example: \textit{(1) physical limitations of space and time could be easily overcome through the metamobility, where we will not move our things, but things will actually move around us; and (2) the future mobility will enable the changes made in the cyber world to be reflected in reality, where cars might become an extension of our own physical senses}.

This article serves as the first effort to offer a comprehensive vision for building up a perpetual, synchronous, and shared metamobility in terms of its system architecture, use cases, and research opportunities. 

\section{Example Architecture of Metamobility}
\label{sc:architecture}
\begin{figure}[t]
    \centering
    \includegraphics[width=\linewidth]{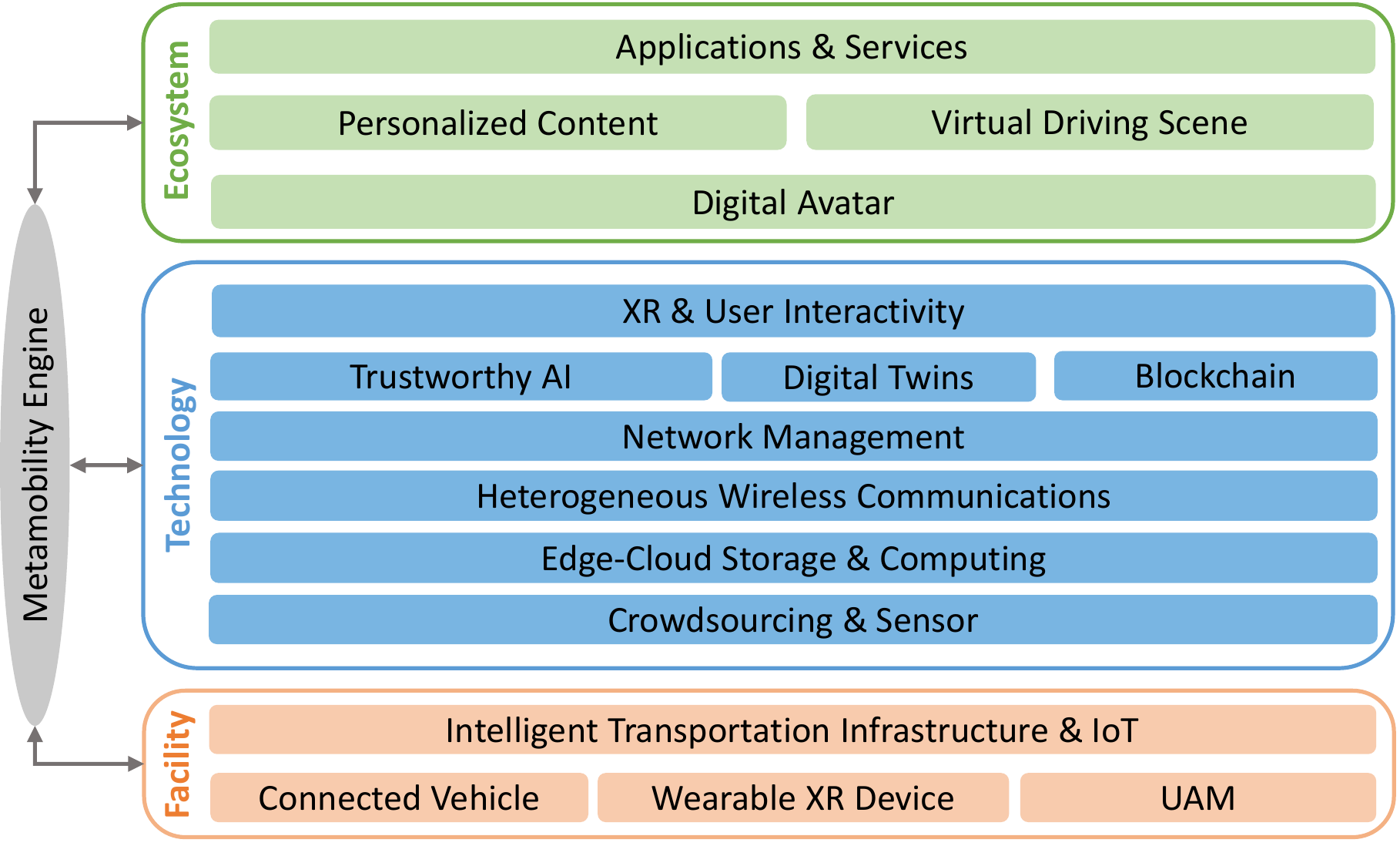}
    \caption{The architecture of the metamobility system.}
    \label{Fig:arch}
\end{figure}

This section presents the basic architecture of the metamobility. It consists of three main parties: facility, technology, and ecosystem, as illustrated in Fig. \ref{Fig:arch}. The facility includes static entities, e.g., intelligent transportation infrastructures, and mobile entities, e.g., CAVs, wearable extended reality (XR) devices, and UAM. These physical entities perform as data generators as well as service requesters. 

From the technology part, we have eight pillars. Physical world data acquisition can be performed in a crowdsourcing manner by leveraging ubiquitous smart sensors deployed in static and mobile entities. When the number of physical entities is sufficiently large, the size of the collected data will be exponentially increased. Edge-cloud storage and computing can relieve the pressure of processing the big data, and enhance the performance of latency-sensitive and computation-intensive mobility applications. To enable the edge-cloud storage and computing, heterogeneous communication technologies, such as 5G, 6G, and cellular vehicle-to-everything (C-V2X), are necessary for supporting high-speed data transmissions. Furthermore, to handle the concurrent data transmission from numerous physical entities, network management strategies, e.g., adaptive data offloading, are required to enhance the data transmission efficiency. AI is delivering on its promise of learning complicated attributes behind the data and performing precise and fast prediction. In the metamobility, trustworthy AI is indispensable due to AI's high involvement in the driving safety. In other words, it is a question of not just what can be done with AI but how it should be done. The digital twin can use the historical data and learned behavior models to conduct scalable simulations in the cyber world and mimic what might happen in the next stage. As these data and digital twin models are usually privacy-concerned, blockchain can be an effective tool to prevent privacy leakage. Last but not least, XR techniques make human users accessible and manipulable to the cyber world.

The ecosystem delineates a perpetual and shared virtual world, a digitized counterpart of the real world. With the assistance of the presented technologies, each physical entity is able to have a unique digital replica or digital avatar. Any changes on the physical entity will result in a real-time digital avatar update accordingly. With continuous input of real world data, the corresponding digital assets, e.g., personalized digital content and virtual driving scene, can be created according to the needs of mobility applications and services.

\section{Example Metamobility Applications}

In this section, we present two metamobility-empowered use cases to demonstrate how the metamobility will benefit and reshape the future mobility systems. Each use case is elaborated from the perspective of the technology evolution, future vision, and key research challenges, respectively.

\subsection{Tactile Live Maps}
\begin{figure*}[t]
    \centering
    \includegraphics[width=\linewidth]{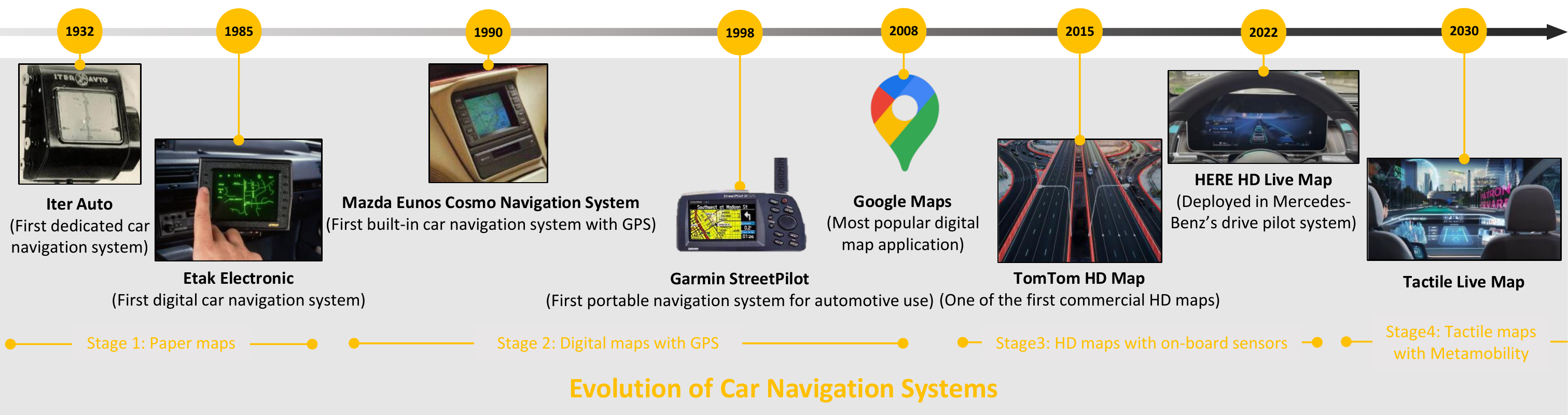}
    \caption{Evolution of car navigation systems from 1932 to 2030, including four stages: paper maps, digital maps with GPS, HD maps with on-board sensors, and tactile live maps with metamobility.}
    \label{Fig:tactilemap}
\end{figure*}

\subsubsection{Evolution}
Car navigation systems have already been shaping the driving experience in an unprecedented way. Most drivers today rely heavily on the modern navigation systems such as Google Maps, which has grown into a multi-billion-dollar industry. But it all started with the Iter Avto, the first dedicated car navigation system created in the 1930s. As illustrated in Fig. \ref{Fig:tactilemap}, it is striking how far we have come with car navigation systems in the last one hundred years. Generally, the evolution of the automotive navigation has passed through three key stages as follows. 
\begin{enumerate}
    \item Paper maps: a set of folded paper maps are shoved in the glove compartment and wrapped from one roll to another across a display. The scroll rate is proportional to the speed of the car.
    \item Digital maps with Global Positioning System (GPS): driven by the commercialization of GPS and miniaturization in electronic devices, vehicle location can be tracked in real-time from satellites in space and mapping information can be stored in data storage. GPS-based navigation has been much more affordable since Google Maps was unveiled.
    \item High-definition (HD) maps with vehicle on-board sensors: as the driving task gradually shifts from the driver to in-vehicle automated systems, the role and scope of digital maps shifts accordingly. As a result, a new generation of maps built purposely for machines is needed. The latest generation of maps, generally referred to as HD maps, comes in the form of a highly accurate and realistic representation of the road. HD maps can be used to help an automated vehicle precisely localize itself on the road (e.g., lane-/centimeter-level localization), understand its surroundings, and plan maneuvers. Although there are some adopted structures for HD maps on the market, such as TomTom \cite{TomTom} and HERE \cite{HERE}, deploying HD maps widely is still in its infancy and considerably costly.
\end{enumerate}
What could be the next item on the agenda?
As listed above, either conventional digital or HD maps mainly target the enhancement of car localization precision. \textit{Could next-generation car navigation systems serve to make driving more accessible, interactable, and entertainable?} Metamobility can play a critical role.

\begin{figure*}[t]
    \centering
    \includegraphics[width=\linewidth]{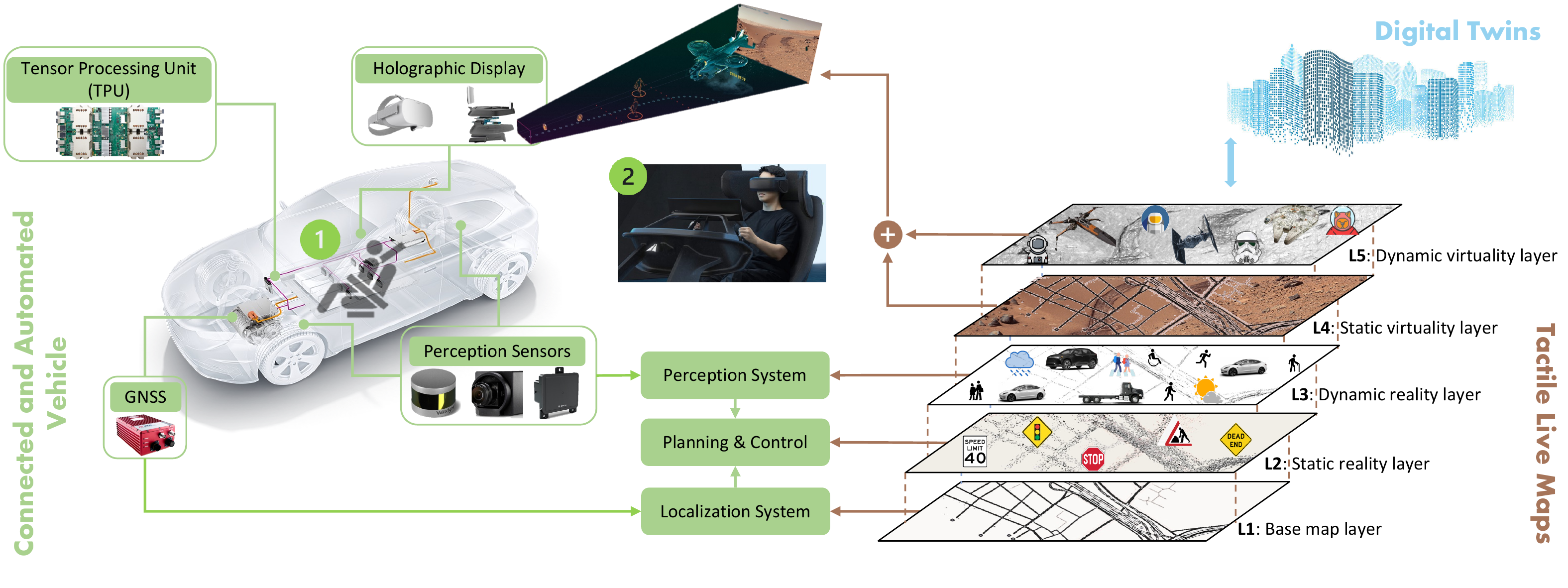}
    \caption{The architecture and eco-system of the proposed tactile live maps.}
    \label{Fig:tlm}
\end{figure*}

\begin{table*}[t]
\centering
\caption{Comparison of digital maps, high definition maps, and tactile live maps}
\begin{tabular}{|l|l|l|l|l|}
\hline
 \multicolumn{2}{|c|}{}& \textbf{Digital Maps w/ GNSS} & \textbf{High Definition Maps} & {\textbf{Tactile Live Maps}} \\\hline
\multicolumn{2}{|c|}{\multirow{2}{*}{\textbf{Service Objects}}} & 
\multirow{2}{*}{Driver} & 
\multirow{2}{*}{Connected and automated vehicle} & Driver\\
\multicolumn{2}{|c|}{} &  &  &  Connected and automated vehicle \\\hline

\multicolumn{2}{|c|}{\multirow{5}{*}{\textbf{Applications}}} & 
Near real-time traffic updates & 
Near real-time traffic updates & Real-time traffic updates \\
\multicolumn{2}{|c|}{} & Eco-friendly routes & On-ramp merges & Immersive navigation \\
\multicolumn{2}{|c|}{} & Street view &  L3-4 automated driving & L5 automated driving \\
\multicolumn{2}{|c|}{} &  &  & Personalized ADAS with XR \\
\multicolumn{2}{|c|}{} &  &  & Interactive in-vehicle gaming \\
\hline

\multicolumn{2}{|c|}{\textbf{Computation Requirements}} & 
Centralized computing & 
Centralized computing &  Hierarchical computing\\
\hline

\multicolumn{2}{|c|}{\multirow{4}{*}{\textbf{Communication Requirements}}} & 
\multirow{2}{*}{Vehicle-to-cloud} & 
\multirow{2}{*}{Vehicle-to-cloud} & Vehicle-to-cloud \\
\multicolumn{2}{|c|}{} &  &  & Vehicle-to-vehicle  \\
\multicolumn{2}{|c|}{} &  &  & Vehicle-to-infrastructure  \\
\multicolumn{2}{|c|}{} &  &  & Vehicle-to-pedestrian  \\
\hline

\multicolumn{2}{|c|}{\multirow{5}{*}{\textbf{Layers}}} & 
Base map layer & Base map layer & Base map layer\\
\multicolumn{2}{|c|}{} & Traffic layer & Geometric map layer &  Static reality layer\\
\multicolumn{2}{|c|}{} &  & Semantic map layer &  Dynamic reality layer\\
\multicolumn{2}{|c|}{} &  & Map priors layer &  Static virtuality layer\\
\multicolumn{2}{|c|}{} &  &  & Dynamic virtuality layer \\
\hline

\multirow{10}{*}{\textbf{KPI}} & End-to-end latency & 1 s & 10 ms & 10-100 $\mu$s \\\cline{2-5}
& Scalability  & Outstanding & Moderate & Challenging \\\cline{2-5}
& \multirow{2}{*}{Accuracy} & Car localization: 5-20 m & Car localization: 5-30 cm & Car localization: 5 cm \\
&  &  &  & Environment perception: very high \\\cline{2-5}
& Data efficiency & $1\times$ & $3\times$ that of digital maps & $5-10\times$ that of HD maps \\\cline{2-5}
& Download data size & $5$ MB/hour& $100$ MB/hour & $1$ GB/hour \\\cline{2-5}
& Upload data size & $\le 1$ MB/day & $50$ MB/day & $2$ GB/day \\\cline{2-5}
& \multirow{3}{*}{Maintenance frequency} & Traffic data: minutes & Traffic data: minutes & Traffic data: real-time \\
&  & Geographic info: days & Geographic info: days & Geographic info: minutes \\
&  & Satellite view: months &  & Virtual content: real-time\\
\hline

\multicolumn{2}{|c|}{\multirow{9}{*}{\textbf{Technologies}}} & 
Cloud computing & Cloud computing & Cloud/edge computing\\
\multicolumn{2}{|c|}{} & GNSS & Over-the-air (OTA) update & Anytime OTA update \\
\multicolumn{2}{|c|}{} & 4G & 5G & 6G \\
\multicolumn{2}{|c|}{} &  & Perception & Cooperative perception\\
\multicolumn{2}{|c|}{} &  &  & Digital twins \\
\multicolumn{2}{|c|}{} &  &  & Metamobility \\
\multicolumn{2}{|c|}{} &  &  & AR/VR \\
\multicolumn{2}{|c|}{} &  &  & Holographic communication \\
\multicolumn{2}{|c|}{} &  &  & mmWave communication \\
\hline

\multicolumn{2}{|c|}{\multirow{8}{*}{\textbf{Vehicle On-board Hardware}}} & On-board navigator &  Camera  & Stereo camera \\
\multicolumn{2}{|c|}{} &  & Inertial measurement unit (IMU) & IMU \\
\multicolumn{2}{|c|}{} &  & Radar & Radar \\
\multicolumn{2}{|c|}{} &  & In-vehicle GPU & In-vehicle TPU \\
\multicolumn{2}{|c|}{} &  &  & Holographic display \\
\multicolumn{2}{|c|}{} &  &  & Wearable XR device \\
\multicolumn{2}{|c|}{} &  &  &  LiDAR\\
\hline

\multicolumn{2}{|c|}{\textbf{Human-Map Interaction}} & Vision, audition & Vision, audition & Vision, audition, tactition, olfaction\\
\hline

\multicolumn{2}{|c|}{\textbf{Data Collection}} & Street view fleet vehicle &  Mobile mapping fleet vehicle & Crowdsourcing\\
\hline

\multicolumn{2}{|c|}{\multirow{3}{*}{\textbf{Map Content}}} & Road geometry, traffic data & Road geometry, lane models, & Road geometry, lane models, traffic data, \\
\multicolumn{2}{|c|}{} &  & traffic data & behavior models of surrounding dynamic \\
\multicolumn{2}{|c|}{} &  & & objects, custom virtual scene, avatars \\
\hline

\multicolumn{2}{|c|}{\multirow{2}{*}{\textbf{Representative Products}}} & 
Google Maps & 
TomTom HD Maps & \multirow{2}{*}{None} \\
\multicolumn{2}{|c|}{} & Apple Maps & HERE HD Maps &   \\\hline

\end{tabular}
\label{tb:tlm}
\end{table*}

\subsubsection{Vision}
The era of autonomous driving is approaching, where driving will be optional for human drivers as full driving automation matures. Nonetheless, this does not signalize a termination of human access to the steering wheel or other driver controls. Today, cars with an automatic transmission still provide a manual mode for drivers who want to shift for themselves. Similarly, OEMs may retain manned driving features in some models to ingratiate themselves with specific customer segments. Then, what are the essential reasons that attract people to drive by themselves? One of the red-hot responses is to savor a more accessible, interactable, and entertainable driving experience, which self-driving alone will never be able to provide. \textit{Tactile live maps, empowered by metamobility, will be the tacit complement to autonomous driving techniques and the key enabler to enhance the accessibility, interactability, and entertainability of CAVs.}

The architecture of the tactile live map ecosystem is illustrated in Fig. \ref{Fig:tlm}, which consists of tactile live maps, digital twins, and CAVs. Specifically, a tactile live map is classified into five layers according to the time intervals at which map information changes (i.e., dynamic or static), and the function of either augmenting the perception of the real environment or enabling the immersive interaction with virtual scenes and content (i.e., reality or virtual). These five layers are perfectly aligned with each other and indexed to allow for efficient parallel information deliveries to corresponding CAV system components, such as the perception system, localization system, planning \& control, and holographic display. Detailed attributes of each layer are:
\begin{itemize}
    \item L1: The base map layer is the bottom layer and contains the basic road network data and $3$D information of the region. It is key for aligning the subsequent layers of the tactile live map such as the static reality layer.
    \item L2: The static reality layer comprises semantic objects that change at intervals of days or hours in reality. Semantic objects include various $2$D and $3$D objects such as stop signs and traffic lights, and positions and state information of road work, accidents, and lane closures. 
    \item L3: The dynamic reality layer includes real-time positions and state information of pedestrians, CAVs, bicycles, motorbikes, etc. It is designed to support the gathering and sharing of real-time global information between a whole fleet of CAVs moving in a specific region. Additionally, with real-time data sampling and historical data provided by digital twins, these dynamic physical objects are endowed with ``souls'' and behaviors, which could serve to conduct timely behavior prediction and anomaly detection.
\end{itemize}   
Layers 1-3 in tactile live maps provide information about the static and dynamic parts of the physical world, and are critical to the autonomous driving systems. They are generated and maintained at significantly high fidelity and there is very little ambiguity about what the ground truth is.
\begin{itemize}
    \item L4: The static virtuality layer offers the capability of customizing the driving scene. For example, a virtual scene, driving on Mars surface, can be chosen by the driver or passengers to enrich the entertainability of mobility. The selected virtual scene will overlay the reality by leveraging XR techniques or deep reality head-up display (HUD) \cite{WayRay}. The semantic information in the static virtuality layer is aligned with that in the static reality layer. 
    
    \item L5: The dynamic virtuality layer is a virtual counterpart of the dynamic reality layer, where each pedestrian or car is represented by its own avatar produced in the metaverse. This layer enhances the accessibility and interactability of the mobility by being linked with the metaverse to provide users with multiple ways, such as through fingers, voice, eyes, and neural signals, to interact with others in real time.
\end{itemize}
Layers $4$ and $5$ are the keys for tactile live maps to evolve from CAV-oriented to both human- and CAV-oriented, and to be with human-like consciousness. The tactile live maps can be utilized by not only drivers physically sitting in cars (i.e., option 1 in Fig. \ref{Fig:tlm}), but also qualified users who are driving remotely with XR remote control systems through holographic communications (i.e., option 2 in Fig. \ref{Fig:tlm}). 

Furthermore, as the dynamic layers are time-sensitive and require real-time maintenance to ensure the data freshness and precision, it is untenable to gather and integrate the offloaded sensor data from CAVs on a city scale. An applicable solution is to segment a city into corridors and to deploy a designated local edge server at each. Hence, tactile live maps could be built and maintained in a scale of the corridor by leveraging edge computing. The key performance indicators (KPIs) for evaluating tactile live maps include end-to-end latency, scalability, accuracy, data efficiency, download/upload data size, and maintenance frequency. Table \ref{tb:tlm} summarizes the KPIs, features, requirements, and technologies of human-oriented digital maps, CAV-oriented HD maps, and everything-oriented tactile live maps.

\subsubsection{Key Research Challenges}
\begin{itemize}
    \item \textbf{Network slicing: } To assure the stringent performance requirement, e.g., latency and reliability, network slicing is necessitated to reserve and isolate particular network resources, e.g., radio bandwidth. As tactile live maps involve complex transmission and computation, their resource demands vary in different technical domains, e.g., radio access, transport, core, and edge networks. Hence, it will be challenging to determine and optimize the resource requests for tactile live maps under varying spatiotemporal network dynamics, e.g., mobility and traffic. As existing network slicing technologies focus on resource reservation at the slice level, rather than users, the achieved performance of individual users may also be different. These differences need to be diminished and even eliminated, which requires further investigation efforts in multiple aspects, e.g., user scheduling and flow controlling.
    \item \textbf{Time-consistent map maintenance:} Maintaining tactile live maps, especially the dynamic layers, replies on a highly dynamic (i.e., CAVs come and go in hard-to-predict manners), large-scale, and decentralized topology. Each CAV may locally hold a self-governing sense-process-offload pipeline, making transforming and merging individual sensor data into a unified live map considerably challenging to be time-consistent. Time inconsistency can compromise the value and precision of real time map information. Hence, there is a need for effective schemes that can overcome the time inconsistency in tactile live map maintenance.

    \item \textbf{Perceivable avatar generation:} Avatars serve as primary digital representatives when we interact with other entities (e.g., human drivers, CAVs in unmanned driving mode, or smart roadside units) in tactile live maps. Thus, both human drivers and CAVs would rely on their avatars to express themselves in the virtual space. For example, a human driver can present his/her current mood through his/her avatar, while a CAV in unmanned driving mode can describe its real-time status in a way that is readily perceivable by humans leveraging its avatar as well, and vice versa. Therefore, research efforts on generating perceivable avatars that can be used in diversified interactions (e.g., human-to-human, human-to-CAV, CAV-to-human, and CAV-to-CAV) are required to accomplish everything-oriented tactile live maps.

\end{itemize}

\begin{figure*}[t]
    \centering
    \includegraphics[width=\linewidth]{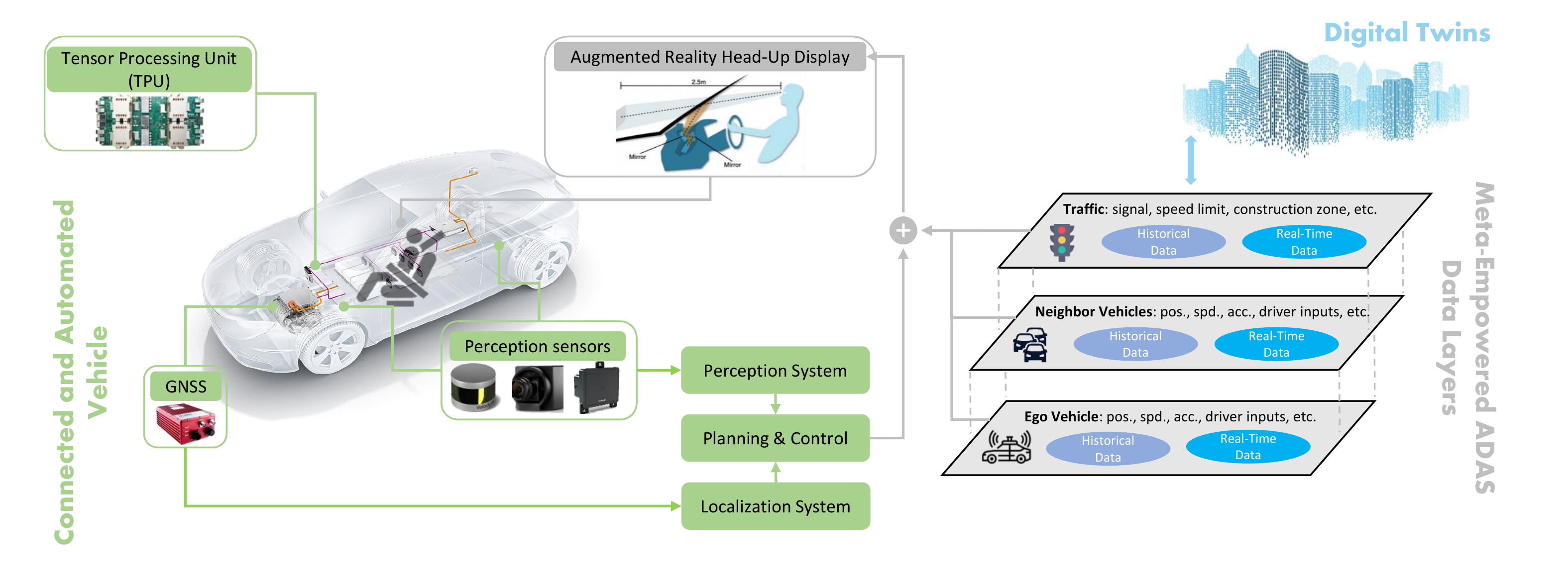}
    \caption{The architecture and eco-system of the meta-empowered ADAS.}
    \label{Fig:ADAS}
\end{figure*}

\subsection{Meta-Empowered ADAS}
\subsubsection{Evolution}
Advanced driver-assistance systems (ADAS) have matured during the last decade, with estimated global market size of USD 25.92 billion in 2021 \cite{ADAS}. They are designed by automotive OEMs and their Tier-1 suppliers to either inform drivers or directly engage vehicle actions in driving and parking scenarios. Since human errors play a big role in traffic accidents, the emergence of ADAS significantly eases the burden on drivers, making driving more relaxed and safer at the same time.

Most commercially available ADAS on the current market rely on the on-board perception of real-time data, and conduct on-board computing based on these perceived data. For example, Adaptive Cruise Control (ACC) systems use the radar equipped on the front bumper of the vehicle (mostly behind the vehicle badge), together with the camera on the windshield, to identify the preceding vehicle and measure its relative speed and distance compared to the ego vehicle. The on-board computer of the vehicle will then calculate its acceleration and braking inputs to adjust its speed and maintain a safe preset distance from the preceding vehicle. Another widely used ADAS on current vehicles is the pre-collision assist system with Automatic Emergency Braking (AEB), which uses the camera on the windshield to continuously detect a potential collision with a vehicle or pedestrian directly ahead of the ego vehicle, and produces visual and audio warning messages to the driver (and apply brakes automatically when necessary).

\subsubsection{Vision}
Although existing ADAS are useful in certain traffic scenarios, they are limited to real-time, short-range information perceived by on-board perception sensors of the ego vehicle. Actions made by ADAS are carried out by vehicle on-board computations, where the behavior-prediction and decision-making processes rely on the perceived data without any access to historical, large-region information. Moreover, existing ADAS always come with a handful of factory settings, which leave very few options for human drivers to customize to satisfy their personalized preferences.

Traditional ADAS can be transformed into meta-empowered ADAS by leveraging emerging technologies, such as XR, edge-cloud storage and computing, and heterogeneous communication technologies (e.g., 5G, 6G, and C-V2X). More importantly, meta-empowered ADAS can take advantage of multiple sources of data beyond the ego vehicle, which are capable of making more informed decisions than simply relying on a single data source.

The architecture of the meta-empowered ADAS is illustrated in Fig. \ref{Fig:ADAS}, where three data layers complement each other by bringing valuable data that can contribute to the construction of digital twins. Each layer contains not only real-time data that can be sampled from hardware sensors on vehicles or traffic infrastructures, but also historical data that are previously sampled and stored for future reference.  

\begin{itemize}
    \item Ego Vehicle Layer: The vehicle dynamics data, such as longitudinal/lateral position, speed, and acceleration, can be sampled in real time through CAN BUS. These data are essential for the meta-empowered ADAS to understand the current status of CAVs and make accurate decisions. Driver inputs can also be sampled from multiple sources, from active inputs (e.g., steering wheel, acceleration/braking pedal, and human-machine interface) to passive inputs (e.g., in-cabin monitoring camera, seat pressure sensor, and steering wheel pressure sensor). Different from vehicle dynamics data, driver inputs provide more information about human factors, which enable the meta-empowered ADAS to be human-centric and can serve humans' needs. Besides the real-time data of vehicle dynamics and driver inputs, their historical data that are also stored in digital twins are also leveraged in this data layer. The access to historical data is a major advantage that allows the meta-empowered ADAS to learn from past performances and preferences of both the vehicle and the driver, hence providing more customized services to meet their needs.
    
    \item Neighbor Vehicle Layer: Different from the aforementioned layer, data in this layer come from neighboring vehicles of the ego vehicle. Getting to know the dynamics of neighboring vehicles is indispensable for most ADAS, as systems like ACC or AEB need to make decisions for the ego vehicle according to the statuses of its neighboring vehicles. Their drivers' statuses can also be helpful in allowing the ego vehicle to predict their future behaviors. These real-time data can be sampled by the ego vehicle through vehicle-to-vehicle (V2V) communications. Moreover, the historical data of neighboring vehicles and their drivers can be sampled by accessing their digital twins, enabling the meta-empowered ADAS to understand the historical trends of neighboring vehicles' statuses and make more informed predictions.  
    
    \item Traffic Layer: This data layer provides the traffic information to the meta-empowered ADAS, such as traffic Signal Phase and Timing (SPaT), speed limits, construction zone alerts, traffic congestion levels, etc. This real-time and historical information allows the ego vehicle to look further down the road instead of only focusing on its surroundings. For example, if the ego vehicle gets the downstream SPaT and congestion data through vehicle-to-infrastructure (V2I) communications, its meta-empowered ADAS can proactively compute an eco-friendly speed trajectory to allow the ego vehicle to pass multiple signalized intersections along that corridor without any full stop at red signals.
\end{itemize}

\begin{figure}[t]
    \centering
    \includegraphics[width=\linewidth]{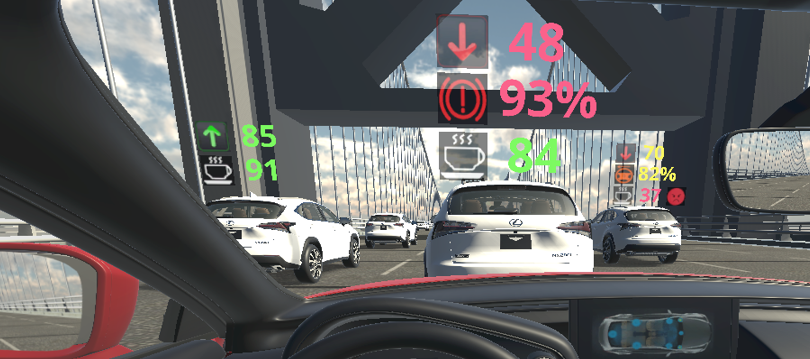}
    \caption{An example illustration of meta-empowered ADAS: Information is displayed on the ego vehicle's windshield as the augmented reality-based head-up display, which includes neighboring drivers' 1) proficiency scores and their trends, 2) possibilities of certain potential actions (e.g., hard braking and lane change), and 3) current mood score.}
    \label{Fig:AR}
\end{figure}

Once aforementioned data are retrieved from these three data layers and digital twins, the planning \& control component of the ego vehicle applies advanced algorithms to process them and generate guidance to the driver through human-machine interfaces. For meta-empowered ADAS, the human-machine interface can be designed as an augmented reality-based head-up display (illustrated in Fig. \ref{Fig:AR}). Guidance information can be visualized to the driver by the projection on the windshield, where neighboring vehicles' driver inputs are overlaid on top of each vehicle. This design outperforms traditional ADAS by enabling the driver to make more informed driving decisions (e.g., decelerate or conduct a lane change to avoid rear-end crashes before the preceding vehicle has a hard braking).

\subsubsection{Research Challenges}
\begin{itemize}
    \item \textbf{Driver distraction: } Meta-empowered ADAS can bring auxiliary sources of information to vehicles, creating an immersive driving experience for drivers. However, since ADAS still require vehicles to be fully or partially driven by human drivers, how to avoid overwhelming drivers with too much information becomes important. Existing ADAS choose to use the dashboard (behind the steering wheel) or multi-media screen (on the center console) to display information for drivers, which requires drivers to move their attention away from the road ahead. Recently, installing big touch screens has become a new norm for new vehicle models, which further distracts drivers, especially for those ADAS that ask drivers to touch multiple times to change their ADAS settings. Augmented reality-based head-up displays (shown in Fig. \ref{Fig:AR}) can be adopted to convey information to drivers more intuitively, but these ADAS should also be carefully designed to clarify differences between real and projected objects.
    
    \item \textbf{Model fidelity: } Big data from various sources can be leveraged by meta-empowered ADAS, where ADAS become capable of making more informed decisions. However, more data does not necessarily mean more accurate models, as some data sources are not as important as others. For example, models cannot treat historical driving data from a month ago the same as that from a day ago, as vehicle performances can change significantly during the past month. Also, if a model completely replicates the driver behavior and satisfies all his/her preferences, such ADAS model might be over-fitted without considering realistic vehicle/traffic constraints. Sufficient analysis of the varied importance of data (e.g., via principal component analysis) should be conducted to ensure the model fidelity, allowing meta-empowered ADAS to better assist drivers instead of getting misled by drivers. 
    
    \item \textbf{Communication heterogeneity: } Since one of the major advantages of meta-empowered ADAS over traditional ADAS is its multiple sources of information, which requires wireless communication technologies to transfer that information between vehicles and edge/cloud. However, existing vehicular wireless communication technologies (e.g., C-V2X) cannot always guarantee network accessibility and low latency when CAVs are traveling at fast speeds or in remote areas. This will generate inaccurate guidance for meta-empowered ADAS, as accurate information cannot be received from the edge/cloud in real time. Therefore, how to construct a heterogeneous wireless communication environment for CAVs with multiple network layers (e.g., vehicle, edge, and cloud) and communication technologies (e.g., C-V2X and DSRC) is crucial to solving communication issues for meta-empowered ADAS, allowing CAVs to maintain a seamless connectivity with low latency and cost.
\end{itemize}

\section{Open Issues for Metamobility Deployment}
In the future development and deployment of the metamobility technology in both academia and industry, together with the involvement of CAVs and digital twins, numerous challenges need to be tackled from the perspectives of both research and engineering.

\subsection{Metamobility Standardization}
The metaverse has gained momentum in multiple domains. Meta has released its social metaverse ecology, while Nvidia announced a new collaboration with BMW on creating future manufacturing solutions by leveraging Omniverse \cite{Omniverse}, Nvidia's metaverse platform. Similar to social and manufacturing, specific standards for future mobility require to be investigated with the help from both academia and industry. However, achieving a consensus amongst diverse sectors (e.g., telecommunication companies, car manufacturers, transportation agencies, and customers), predictably, might be arduous.

\subsection{Edge AI}
Edge AI, the confluence of edge computing and AI, will be the base to support various features in the metamobility, such as autonomy of avatar, data interoperability, scene understanding, and distributed learning. On the other hand, the main characteristics of the metamobility - the immeasurable source of sophisticated data and high user engagement would provide both challenges and opportunities for AI techniques to achieve efficient data processing, analysis, and training.

\subsection{Network}
The metamobility applications require high throughput (e.g., to upload multi-fold on-board sensor data in real-time), ultra-low motion-to-photon latency (i.e., delay between a user's action and its corresponding reaction on display), and pervasive network access, while physical objects such as CAVs are moving. For example, an unsatisfied motion-to-photon latency might cause car sickness, and thus degrade metamobility experiences. To tackle these issues, two essential techniques should be studied and implemented in the metamobility: (1) context-aware data offloading to adaptively adjust the data collection (i.e., spatial-temporal), perception, and transmission with high mobility; and (2) application-oriented network resource provisioning to reduce cross-domain resource usage while meeting the strict latency requirement.

\subsection{Security and Privacy}
In the metamobility, security researchers have to design new mechanisms to protect legitimate entities, such as physical CAVs or their corresponding digital assets, against attacks from both physical and digital spaces. Compared to the physical entity, its digital asset would be more vulnerable as it usually contains sufficient information such as the driver's biometric data to mimic or even clone the entity. Furthermore, due to the nature of the metamobility, any entity can monitor others' activities in the digital space (e.g., layer 5 in tactile live maps). Numerous records of behaviors, user interaction traces, and digital replicas will dwell in the metamobility. Hence, how to prevent eavesdropping, continuous monitoring, and privacy leakages is the key to secure the metamobility natives.

\section{Conclusion}

In this article, the ``metamobility'' was first coined and defined to connect future mobility systems with the metaverse. The breakthrough lying before us is to create a frictionless, fun, and personalized mobility society by enabling the metamobility. We illustrated an example architecture of the metamobility that integrates key technologies and facilities to enable a plurality of new mobility applications, products, and services. The exploration of the metamobility will drive innovations in future mobility technologies, industries, and economies, which, in turn, will help to make the world a better place to live.

\bibliographystyle{IEEEtran}
\bibliography{reference}

\begin{thebibliography}{10}
\providecommand{\url}[1]{#1}
\csname url@samestyle\endcsname
\providecommand{\newblock}{\relax}
\providecommand{\bibinfo}[2]{#2}
\providecommand{\BIBentrySTDinterwordspacing}{\spaceskip=0pt\relax}
\providecommand{\BIBentryALTinterwordstretchfactor}{4}
\providecommand{\BIBentryALTinterwordspacing}{\spaceskip=\fontdimen2\font plus
\BIBentryALTinterwordstretchfactor\fontdimen3\font minus
  \fontdimen4\font\relax}
\providecommand{\BIBforeignlanguage}[2]{{%
\expandafter\ifx\csname l@#1\endcsname\relax
\typeout{** WARNING: IEEEtran.bst: No hyphenation pattern has been}%
\typeout{** loaded for the language `#1'. Using the pattern for}%
\typeout{** the default language instead.}%
\else
\language=\csname l@#1\endcsname
\fi
#2}}
\providecommand{\BIBdecl}{\relax}
\BIBdecl

\bibitem{ToyotaCASE}
\BIBentryALTinterwordspacing
Toyota. (2019) Reforming our company to become a ``mobility company". Accessed
  on Sep. 2022. [Online]. Available:
  \url{https://global.toyota/en/company/messages-from-executives/details/}
\BIBentrySTDinterwordspacing

\bibitem{stephenson2003snow}
N.~Stephenson, \emph{Snow crash: A novel}.\hskip 1em plus 0.5em minus
  0.4em\relax Spectra, 2003.

\bibitem{duan2021metaverse}
H.~Duan, J.~Li, S.~Fan, Z.~Lin, X.~Wu, and W.~Cai, ``Metaverse for social good:
  A university campus prototype,'' in \emph{Proc. the 29th ACM International
  Conference on Multimedia}, 2021, pp. 153--161.

\bibitem{lee2021all}
L.-H. Lee, T.~Braud, P.~Zhou, L.~Wang, D.~Xu, Z.~Lin, A.~Kumar, C.~Bermejo, and
  P.~Hui, ``All one needs to know about metaverse: A complete survey on
  technological singularity, virtual ecosystem, and research agenda,''
  \emph{arXiv preprint arXiv:2110.05352}, 2021.

\bibitem{MetaHorizon}
\BIBentryALTinterwordspacing
Meta. (2021) Meta horizon worlds. Accessed on Sep. 2022. [Online]. Available:
  \url{https://www.oculus.com/horizon-worlds/}
\BIBentrySTDinterwordspacing

\bibitem{Omniverse}
\BIBentryALTinterwordspacing
Nvidia. (2021) Nvidia, {BMW} blend reality, virtual worlds to demonstrate
  factory of the future. Accessed on Sep. 2022. [Online]. Available:
  \url{https://blogs.nvidia.com/blog/2021/04/13/nvidia-bmw-factory-future/}
\BIBentrySTDinterwordspacing

\bibitem{jin2017towards}
Y.~Jin, J.~Zhang, M.~Li, Y.~Tian, H.~Zhu, and Z.~Fang, ``Towards the automatic
  anime characters creation with generative adversarial networks,'' \emph{arXiv
  preprint arXiv:1708.05509}, 2017.

\bibitem{zhu2020haptic}
M.~Zhu, Z.~Sun, Z.~Zhang, Q.~Shi, T.~He, H.~Liu, T.~Chen, and C.~Lee,
  ``Haptic-feedback smart glove as a creative human-machine interface ({HMI})
  for virtual/augmented reality applications,'' \emph{Science Advances},
  vol.~6, no.~19, p. eaaz8693, 2020.

\bibitem{clemm2020toward}
A.~Clemm, M.~T. Vega, H.~K. Ravuri, T.~Wauters, and F.~De~Turck, ``Toward truly
  immersive holographic-type communication: Challenges and solutions,''
  \emph{IEEE Communications Magazine}, vol.~58, no.~1, pp. 93--99, 2020.

\bibitem{9724183}
Z.~Wang, R.~Gupta, K.~Han, H.~Wang, A.~Ganlath, N.~Ammar, and P.~Tiwari,
  ``Mobility digital twin: Concept, architecture, case study, and future
  challenges,'' \emph{IEEE Internet of Things Journal}, vol.~9, no.~18, pp.
  17\,452--17\,467, 2022.

\bibitem{taheri2022goal}
O.~Taheri, V.~Choutas, M.~J. Black, and D.~Tzionas, ``{GOAL}: Generating 4{D}
  whole-body motion for hand-object grasping,'' in \emph{Proc. the IEEE/CVF
  Conference on Computer Vision and Pattern Recognition (CVPR)}, 2022, pp.
  13\,263--13\,273.

\bibitem{TomTom}
\BIBentryALTinterwordspacing
TomTom. (2022) Tom{T}om {HD} maps. Accessed on Sep. 2022. [Online]. Available:
  \url{https://www.tomtom.com/en_us/}
\BIBentrySTDinterwordspacing

\bibitem{HERE}
\BIBentryALTinterwordspacing
HERE. (2022) {HERE} {HD} live maps. Accessed on Sep. 2022. [Online]. Available:
  \url{https://www.here.com/platform/HD-live-map}
\BIBentrySTDinterwordspacing

\bibitem{WayRay}
\BIBentryALTinterwordspacing
WayRay. (2022) Deep reality display. Accessed on Sep. 2022. [Online].
  Available: \url{https://wayray.com/deep-reality-display/#about}
\BIBentrySTDinterwordspacing

\bibitem{ADAS}
\BIBentryALTinterwordspacing
V.~M. Research. (2022) Advanced driver assistance systems ({ADAS}) market size
  and forecast. Accessed on Sep. 2022. [Online]. Available:
  \url{https://www.verifiedmarketresearch.com/product/global-advanced-driver-assistance-systems-adas-market-size-and-forecast-to-2025/}
\BIBentrySTDinterwordspacing

\end{thebibliography}

\newpage
\begin{IEEEbiography}
[{\includegraphics[width=1in,height=1.25in,clip,keepaspectratio]{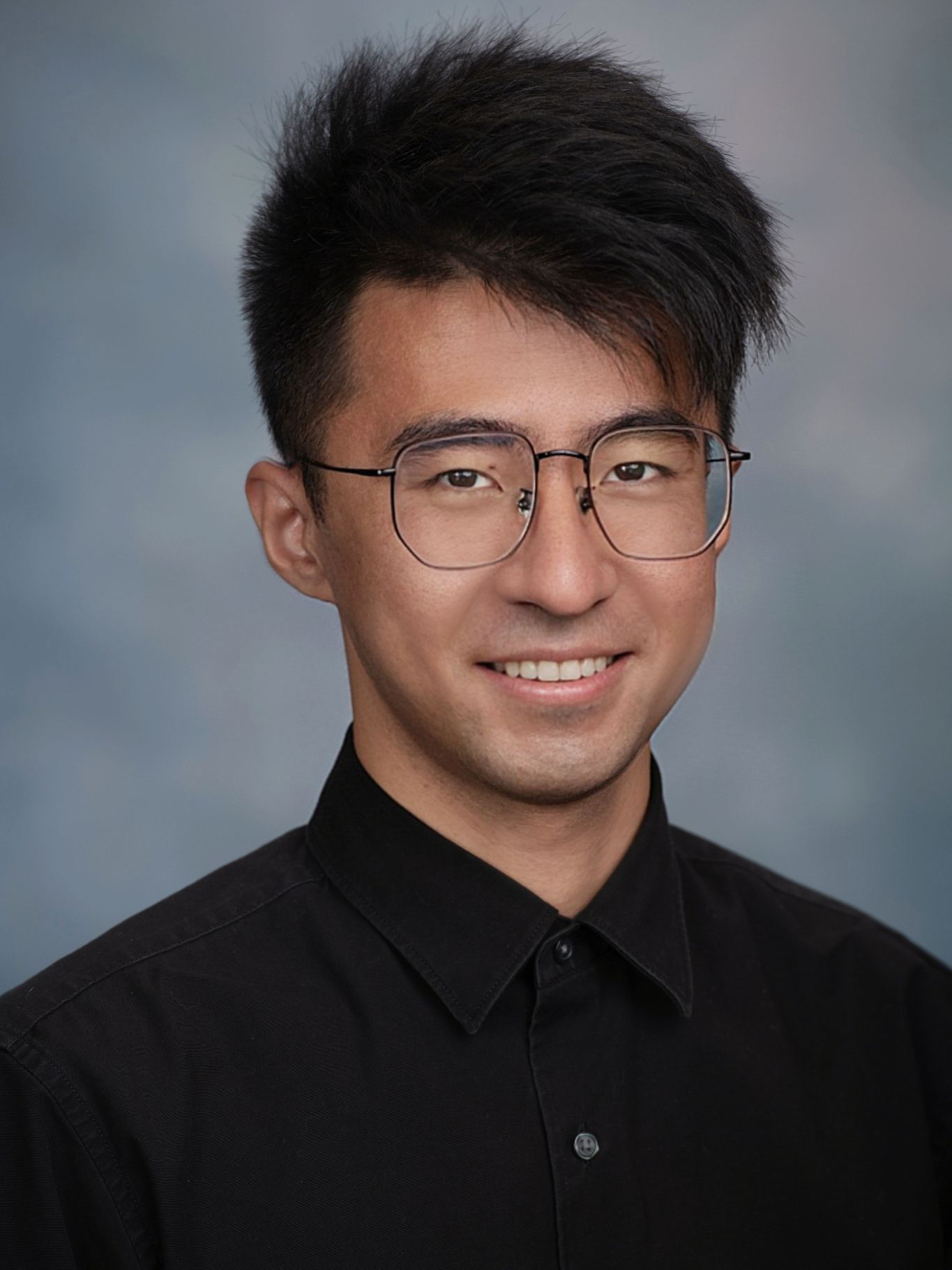}}]
{Haoxin Wang}
(S'16-M'21) is an Assistant Professor in the Department of Computer Science at Georgia State University, and leads the Advanced Mobility \& Augmented Intelligence (AMAI) Lab. He received the Ph.D. degree from The University of North Carolina at Charlotte in 2020. From 2020 to 2022, he was a Research Scientist at Toyota Motor North America, InfoTech Labs. His current research interests include mobile AR/VR, autonomous driving, holographic communications, and edge intelligence.
\end{IEEEbiography}

\begin{IEEEbiography}
[{\includegraphics[width=1in,height=1.25in,clip,keepaspectratio]{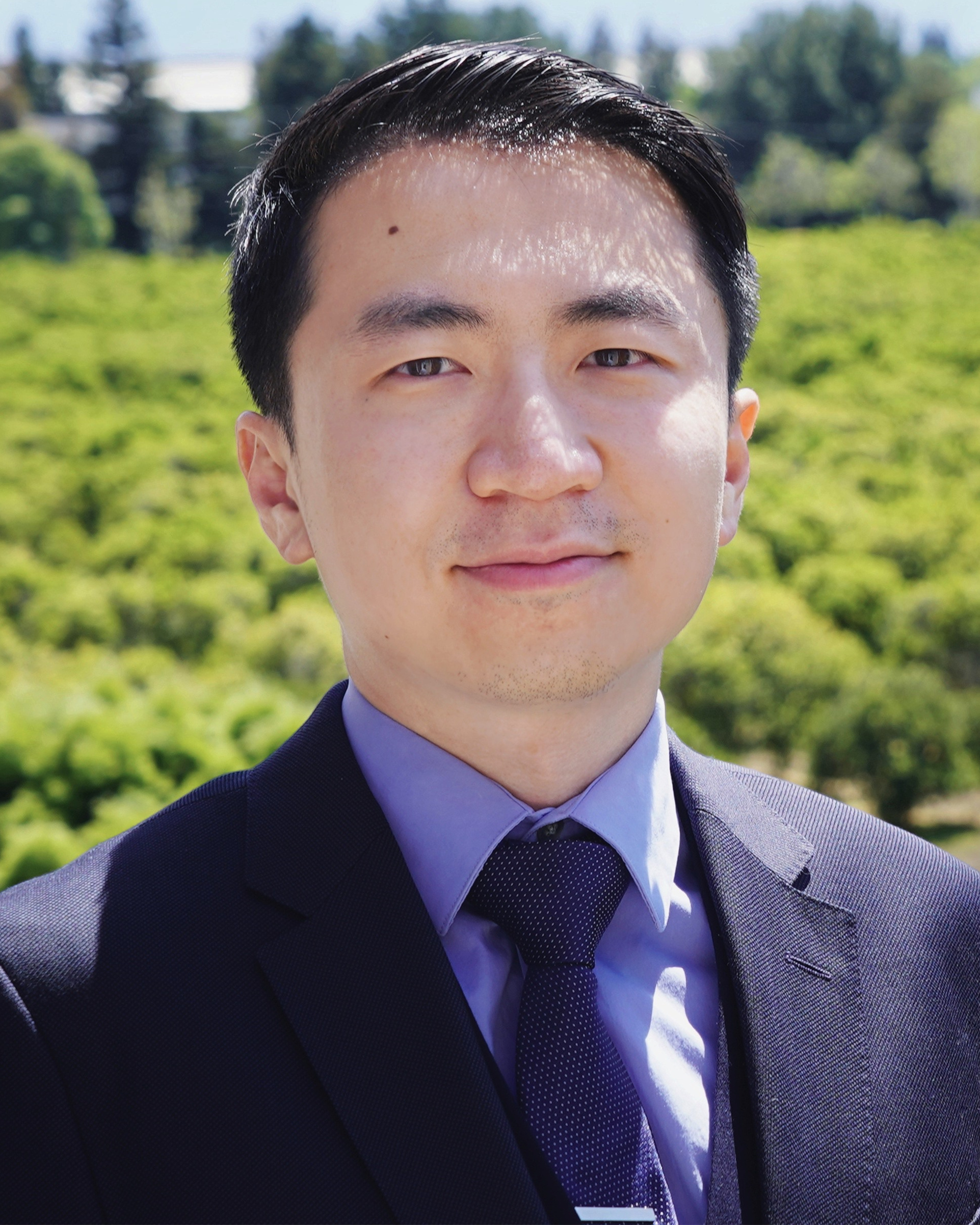}}]
{Ziran Wang}
(S'16-M'19) received the Ph.D. degree from The University of California, Riverside in 2019. He is an Assistant Professor in the College of Engineering at Purdue University, and was a Principal Researcher at Toyota Motor North America. He serves as Founding Chair of IEEE Technical Committee on Internet of Things in Intelligent Transportation Systems, and Associate/Guest Editor of five academic journals. His research focuses on automated driving, human-autonomy teaming, and digital twin.
\end{IEEEbiography}

\begin{IEEEbiography}
[{\includegraphics[width=1in,height=1.25in,clip,keepaspectratio]{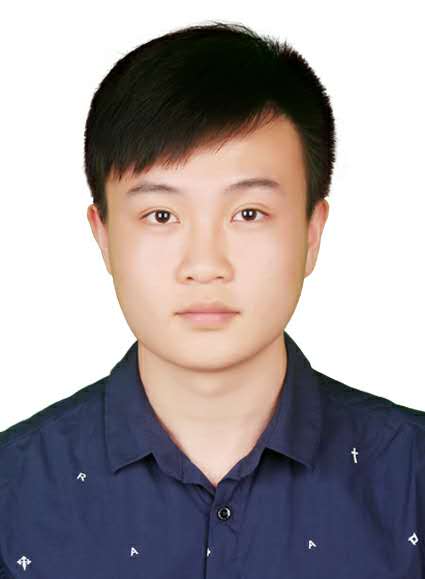}}]
{Dawei Chen}
(S'19-M'22) received the B.S. degree in telecommunication engineering from Huazhong University of Science and Technology, Wuhan, China, and the Ph.D. degree in electrical and computer engineering from the University of Houston, Houston, TX, in 2015 and 2021, respectively. From 2015 to 2016, he was a network engineer at Ericsson. Currently he is a research scientist at Toyota Motor North America, InfoTech Labs. His research interests include deep learning, edge/cloud computing, federated learning/analytics, and wireless networks.
\end{IEEEbiography}

\begin{IEEEbiography}
[{\includegraphics[width=1in,height=1.25in,clip,keepaspectratio]{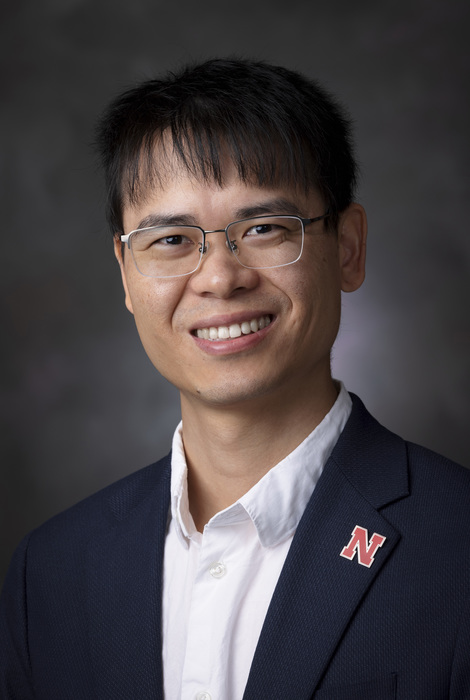}}]
{Qiang Liu}
(S'16-M'21) is an Assistant Professor at the School of Computing, University of Nebraska-Lincoln. He earned his Ph.D. degree in Electrical Engineering from the University of North Carolina at Charlotte (UNCC) in 2020. His papers won IEEE Communications Society's TAOS Best Paper Award 2019, and IEEE International Conference on Communications (ICC) Best Paper Award 2019 and 2022. His research interests lie in the broad field of edge computing, wireless communication, computer networking, and machine learning.  
\end{IEEEbiography}

\begin{IEEEbiography}
[{\includegraphics[width=1in,height=1.25in,clip,keepaspectratio]{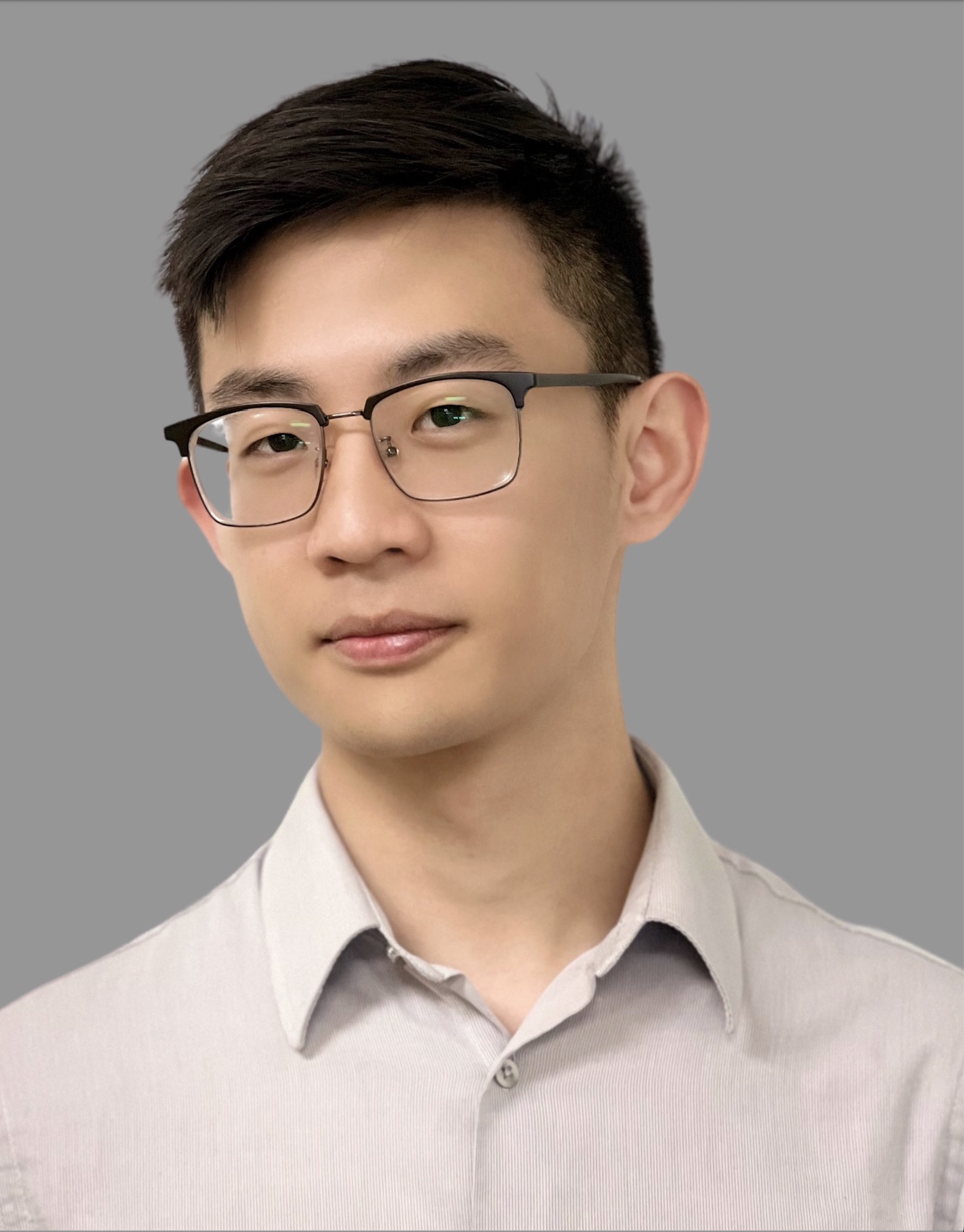}}]
{Hongyu Ke}
received the Master degree from The University of Buffalo in 2022, and Bachelor degree from StonyBrook University in 2021. He is pursing his Ph.D. degree in the Department of Computer Science at Georgia State University. His current research areas include mobile augmented reality, efficient AI, and edge computing.

\end{IEEEbiography}

\begin{IEEEbiography}
[{\includegraphics[width=1in,height=1.25in,clip,keepaspectratio]{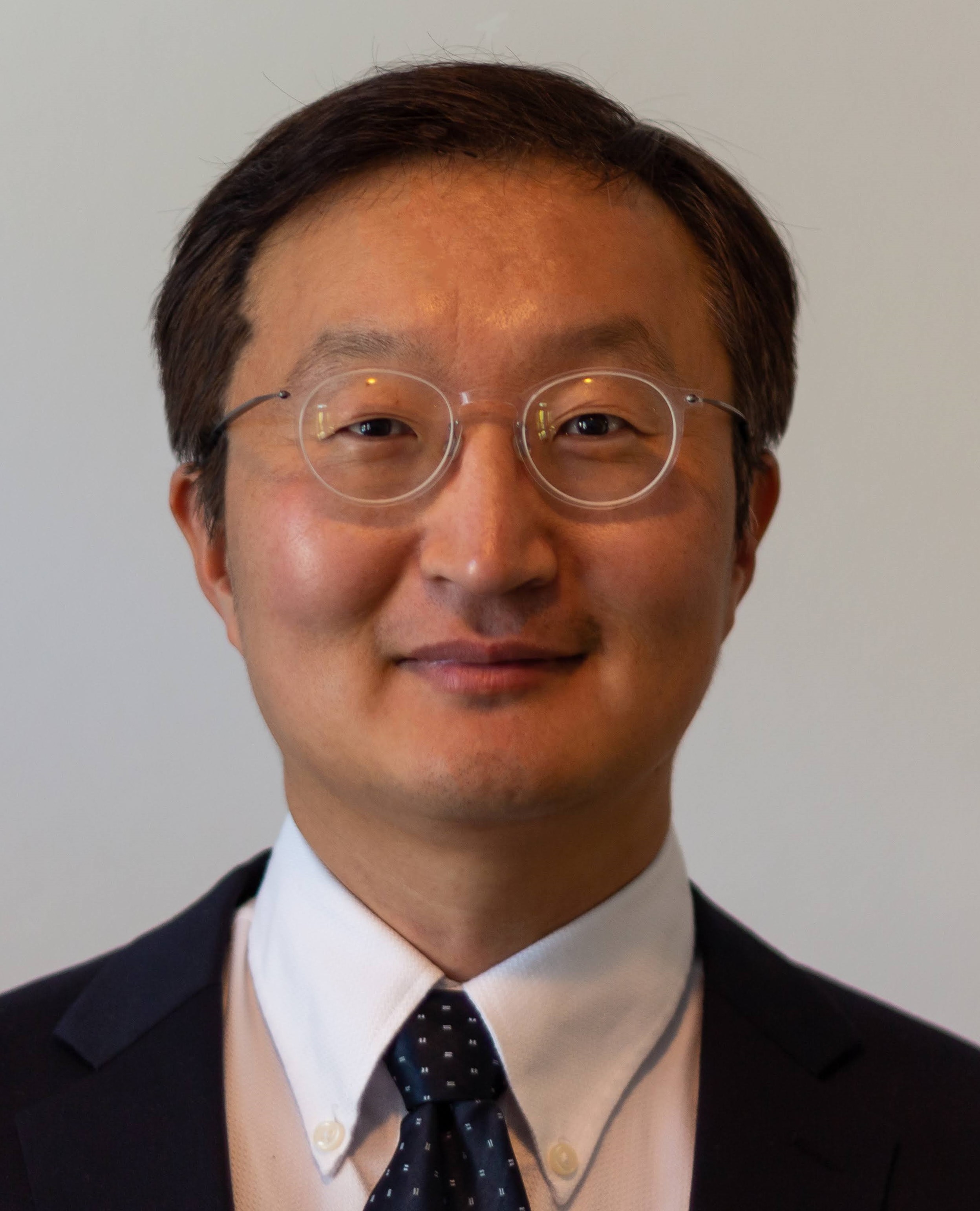}}]
{Kyungtae (KT) Han}
(M'97-SM'15) received the Ph.D. degree in electrical and computer engineering from The University of Texas at Austin in 2006. He is currently a Principal Researcher at Toyota Motor North America, InfoTech Labs. Prior to joining Toyota, Dr. Han was a Research Scientist at Intel Labs, and a Director in Locix Inc. His research interests include cyber-physical systems, connected and automated vehicle techniques, and intelligent transportation systems.  
\end{IEEEbiography}

\end{document}